\documentclass[12pt]{article}
\usepackage{graphicx}
\usepackage{psfrag}

\begin{document}

\begin{titlepage}
\null\vspace{-62pt}

\pagestyle{empty}
\begin{center}

\vspace{1.0truein} {\Large\bf Effective potential for Lifshitz type z=3 gauge theories}

\vspace{1in}
{\large K.~Farakos} \\
\vskip .4in
{\it Department of Physics,\\
National Technical University of Athens,\\
Zografou Campus, 15780 Athens, Greece\\
kfarakos@central.ntua.gr}\\

\vspace{0.5in}

\vspace{.5in}
\centerline{\bf Abstract}

\baselineskip 18pt
\end{center}

\noindent
We consider the one-loop effective potential at zero temperature in Lifshitz type field theories with anisotropic space-time scaling, with critical exponent $z=3$, including scalar, fermion  and gauge fields. The fermion determinant generates a symmetry breaking term at one loop in the effective potential and a local minimum appears, for non zero scalar field, for every value of the Yukawa coupling.
Depending on the relative strength of the coupling constants for the scalar and the gauge field, we find a second symmetry breaking local minimum in the effective potential for a bigger value of the scalar field.

\end{titlepage}

\newpage
\pagestyle{plain}
\setcounter{page}{1}
\newpage

\section{Introduction}

In this work we study four dimensional U(1) gauge theories in the Lifshitz context, interacting with a complex scalar field and fermions with Yukawa coupling to the scalar field.
The discussion on Lifshitz-type models has been stimulated recently by Horava who proposed a
theory of gravity which is power counting renormalizable \cite{hor1}. Extensive research
in the gravitational and cosmological aspects, such as black hole solutions, has been done \cite{kk}.

In the Lifshitz-type models there is an anisotropic scaling between temporal and spatial directions,
measured by the dynamical critical exponent, $z$,
\begin{equation}
t\rightarrow b^z t,\,\,\,x_i\rightarrow b x_i
\end{equation}
that breaks Lorentz symmetry in the classical level.
Non-relativistic field theories of this type have been considered also since they have an improved ultraviolet behavior and their renormalizability properties are quite different than conventional Lorentz symmetric theories \cite{visser}--\cite{jarev}.
Various field theoretical models at the Lifshitz point and non Lorentz invariant extensions of gauge field theories are under consideration \cite{hor3}.  We observe that the quantum effects define an effective energy scale, dependent on the dynamics of the model and restore Lorentz invariance for energies smaller than this scale \cite{iengo, yukawa}.

In order to investigate the various implications of a field theory it is particularly important to examine its symmetry structure, at the classical and the quantum level. A basic tool for this
study is the effective action and effective potential \cite{col}--\cite{farias}.
In a previous work \cite{before} we analyzed, for dynamical exponent $z=2$ the case of a single scalar field and we found a symmetry breaking term induced at one-loop at zero temperature, as well as terms induced at finite temperature at one-loop, that provide symmetry restoration through a first-order phase transition. In a subsequent work \cite{gaugez2} we studied a scalar field coupled with a U(1) gauge field also for $z=2$. We found that the results depend on the relative strength of the coupling constants for the scalar and the gauge fields, for a particular range of values of which we found similar effects as in \cite{before}; namely, when the strength of the gauge interaction is much smaller than that of the scalar the model has spontaneous symmetry breaking effects. The phenomenon of symmetry breaking however disappears for increasing values of the gauge coupling. In high temperature the model shows symmetry restoration effects.

Because of the importance of symmetry breaking and restoration phenomena in particle physics and cosmology we extend here our previous work, including scalar, gauge fields and fermions at the value $z=3$ of the critical exponent\footnote{One cannot define the analog of the Dirac equation for $z=2$.}. Considering the physical implications of the several terms induced at one-loop in the effective potential we see that the behavior of the model depends strongly on the gauge and the fermion couplings with a rich phase structure. The behavior of the terms appearing at one loop from the integration of the various fields differs drastically from what happens correspondingly in the case $z=2$, as we describe in the main text. Depending on the relative strength of the gauge coupling to the scalar coupling one or two minima appear in the effective potential. If the fermion Yukawa coupling is non zero then both minima correspond to symmetry breaking with the relative depth depending on the couplings of the model. We can see that, in this model, fermions can give symmetry breaking without
the need of a coupling to a gauge field \cite{cahill, yukawa}.

We write the model and calculate the
one-loop effective potential in Sec.~2. In Sec.~3 we describe the contribution of the various terms in the
effective potential.

\section{The model, calculation of the effective potential}

The action of the model is
\begin{eqnarray}
S=\int dt d^3x  \biggl(
-\frac{1}{2} F_{0i}F^{0i} -\frac{1}{4}F_{ij}(-\Delta)^{2} F^{ij}
+\frac{1}{2}\left[ D_0\Phi\right]^{\dag}\left[ D_0\Phi\right] \nonumber\\
+\frac{1}{2}\left[D_iD^2\Phi\right]^{\dag}\left[D^iD^2\Phi\right]
+\overline{\Psi}i\gamma^0D_0\Psi+\overline{\Psi}i\gamma^kD_kD^2\Psi
-h\Phi^{\dag}\Phi\overline{\Psi}\Psi-U(\Phi) \biggr)
\end{eqnarray}
with $z=3$ where, as usual, $F_{\mu\nu}=\partial_{\mu}A_{\nu}-\partial_{\nu}A_{\mu}$,
$\Delta=\partial_i \partial_i$, $\Phi=\Phi_1+i\Phi_2$,
$D_{\mu}=\partial_{\mu}+ieA_{\mu}$ and $D^2=D_kD^k=-D_kD_k$.

The dimensions in units of mass for the gauge fields and coupling are: $[A_0]=(d+z-2)/2$, $[A_i]=(d-z)/2$ and $[e]=(z-d)/2+1$ and for case $d=3$, $z=3$ we have $[A_0]=2$, $[A_i]=0$ and $[e]=1$. For the fermion field and
the Yukawa coupling of fermions we have $[\Psi]=d/2$, $[h]=z$.
The scalar field has also dimensionality
$[\Phi]=(d-z)/2=0$,
the potential term for $\Phi$ can, therefore, be an arbitrary polynomial up to the infinite order \cite{Liouville}.

In order to illustrate the general features of the relative contributions of the gauge, scalar and fermion terms to the effective potential we will consider a potential term of the eighth order,
\begin{equation}
U(\Phi)=\frac{g}{8!}\left(\Phi^{\dag}\Phi\right)^4
\label{ss1}
\end{equation}
with $[g]=6$, where the strength of relative contributions, as it will turn out, will depend on the
dimensionless ratio $g/e^6$.

The action is gauge invariant under the usual $U(1)$ gauge transformations.
To calculate the effective potential we add an appropriate, for $z=3$ gauge theories, gauge-fixing term \cite{gaugez2}
\begin{equation}
L_{gf}=-\frac{1}{2}(\partial_0 A_0 +\Delta^2\partial_i A^i)\frac{1}{\Delta^2}
(\partial_0 A_0 +\Delta^2\partial_i A^i).
\label{gft}
\end{equation}
Although non-local, this is the analog of the Feynman gauge for the Lifshitz-type theory at hand and can be derived using either a Fadeev-Popov \cite{gaugez2} or a BRST \cite{ans} procedure. As in usual QED the ghosts decouple and we can ignore them.
This gauge fixing term, (\ref{gft}), has the advantage of making the calculation of the effective action easier by canceling the
mixed $A_0-A_i$ terms.
One may investigate these problems in various, more general gauge conditions; provided one chooses a renormalizable gauge-fixing, the physical results will be gauge-independent \cite{jackiw, dolan, nielsen}.

Now we shift the fields $\Phi_1\rightarrow\phi+\phi_1$,
$\Phi_2\rightarrow 0+\phi_2$ in the action, expand the action up to quadratic order, and then calculate the effective potential for $\phi$ at one-loop integrating over the various fields \cite{jackiw}.
To calculate only the effective potential we keep $\phi$ space-time independent.
We write for the Lagrangian, separating the kinetic terms from the interaction terms:
\begin{equation}
L+L_{gf}=L_2 + L_{int}
\end{equation}
keeping only quadratic order and $\phi$-dependent terms, where
\begin{eqnarray}
L_2= \frac{1}{2}A_0 (\frac{\partial_0^2}{\Delta^2}-\Delta+m_e^2(\phi)) A_0
- \frac{1}{2}A_i (\partial_0^2-\Delta^3) A_i \nonumber\\
-\frac{1}{2}\phi_1 (\partial_0^2-\Delta^3+m_1^2(\phi)) \phi_1
-\frac{1}{2} \phi_2 (\partial_0^2-\Delta^3+m_2^2(\phi)) \phi_2 \nonumber\\
+\overline{\Psi}(i\gamma^0 \partial_0-i\gamma^k \partial_k \Delta-m_F(\phi))\Psi
\end{eqnarray}
and
\begin{equation}
L_{int}= m_e(\phi) A_0 \partial_0\phi_2
-\frac{1}{2}m_e(\phi)^2 A_i \partial_i \partial_j \Delta A_j
- m_e(\phi) A_i \partial_i \Delta^2 \phi_2.
\end{equation}
We have $m_1^2(\phi)=U''(\phi)$,
$\phi m_2^2(\phi) = U'(\phi)$ for the scalar
$\phi$-dependent masses. For the fermion and the gauge field we have correspondingly
$m_F(\phi)=h \phi^2$ and $m_e(\phi)=e \phi$. For the particular potential term that we will analyze we have
$m_1^2(\phi) =\frac{1}{6!}g \phi^6$ and $m_2^2(\phi)= \frac{1}{7!} g \phi^6$, but we will keep the abstract expressions in order to give the most general result for the effective potential.\\

If we turn now to the Euclidean momentum space after the Wick rotation, the action of the model is:
\begin{eqnarray}
S_E=\frac{1}{2}\int \frac{d\omega d^3 p}{(2\pi)^4} \{A_4(\frac{\omega^2}{p^4}+p^2+m_e^2) A_4
+A_k(\omega^2+p^6)A_k
+\phi_1(\omega^2+p^6+m_1^2)\phi_1 \nonumber\\
+\phi_2(\omega^2+p^6+m_2^2)\phi_2 + m_e^2 A_k p_k p_j p^2 A_j
-2im_e A_4 \omega \phi_2 + 2im_e A_k p_k p^4 \phi_2 \} \nonumber\\
+\int \frac{d\omega d^3 p}{(2\pi)^4}\overline{\Psi}(\gamma_4 \omega+\gamma_k p_k p^2 - m_F)\Psi,~~~~
\end{eqnarray}
where we used the expressions $\omega=ip_0$, $\gamma_4=i\gamma_0$, $A_0=iA_4$ and the notation: $p^2 =p_i^2$ for the spatial momentum. We use also the abbreviations $m_k$ instead of $m_k(\phi)$ in what follows.

When we are doing the functional integration it is easier to perform
first the integration with respect to $A_4$ and $A_k$, and then with respect to $\phi_2$.
The integration with respect to $\phi_1$ and $\overline{\Psi},\Psi$ is independent and straightforward.

The standard techniques of functional integration \cite{jackiw} give the final result for the effective potential at one-loop:
\begin{eqnarray}
U_{\rm eff}= U(\phi)+\frac{1}{2}Tr\ln(\omega^2+p^6+m_1^2) + \frac{1}{2}Tr\ln(\omega^2+p^6+m_2^2) \nonumber\\
-2Tr\ln(\omega^2+p^6+m_F^2)
+ \frac{1}{2}Tr\ln(\omega^2+p^6+m_e^2p^4) \nonumber\\
+\frac{1}{2}Tr\ln\left(1+\frac{m_e^2 m_2^2 p^4}{(\omega^2+p^6)(\omega^2+p^6+m_2^2)}\right),
\label{effpot}
\end{eqnarray}
where $Tr=\int\frac{d\omega d^3 p}{(2\pi)^4}$.

The first three contributions in (\ref{effpot}) can be treated in the same way.
For the integration over $\omega$ after dropping an overall infinite constant we have \cite{gaugez2, farias}:
\begin{equation}
\int_{-\infty}^{\infty} d\omega \ln(\omega^2 + B)=2\pi\sqrt{B}.
\end{equation}
We use also the regularized identity:
\begin{equation}
\int_0^\infty d\alpha \alpha^{-3/2-\delta}\exp^{-\alpha B}= \Gamma(-\frac{1}{2}-\delta)B^{\frac{1}{2}+\delta}=-2\sqrt{\pi}\sqrt{B},
\end{equation}
in the limit $\delta \rightarrow 0$.

So the first one-loop correction term, $U_{m_1}$, in (\ref{effpot}) becomes
\begin{eqnarray}
U_{m_1}=\frac{1}{2}\int\frac{d\omega d^3 p}{(2\pi)^4}\ln(\omega^2+p^6+m_1^2)
=\frac{1}{2}\int\frac{d^3 p}{(2\pi)^3}\sqrt{p^6+m_1^2} \nonumber\\
=\frac{M^\epsilon}{2}\int\frac{d^d p}{(2\pi)^d}\sqrt{p^6+m_1^2}
=-\frac{M^\epsilon}{4\sqrt{\pi}}\int_0^\infty d\alpha \alpha^{-3/2}\int\frac{d^d p}{(2\pi)^d}
\exp^{-\alpha(p^6+m_1^2)},
\end{eqnarray}
where $\epsilon=3-d$ and $M$ is an arbitrary UV mass scale introduced to compensate for the integral
dimensions, keeping $[p]=1$.
Integrating out the angles of the integration momentum and after the change of variable $p=u^{\frac{1}{3}}$
we get
\begin{equation}
U_{m_1}=-\frac{M^\epsilon}{4\sqrt{\pi}}\frac{1}{(2\pi)^d}\frac{2 \pi^{d/2}}{\Gamma(d/2)}
\int_0^\infty d\alpha \alpha^{-3/2}\int_0^\infty \frac{du}{3} u^{\frac{d}{3}-1}
\exp^{-\alpha(u^2+m_1^2)}.
\end{equation}
We perform the integral over $u$ first and then the integral over $\alpha$, and using a change of variables and the properties of the $\Gamma$ function, we get:
\begin{eqnarray}
U_{m_1}=-\frac{M^\epsilon}{4\sqrt{\pi}}\frac{1}{(2\pi)^d}\frac{2 \pi^{d/2}}{\Gamma(d/2)}
\frac{\Gamma(d/6)}{6} \Gamma(-\frac{1}{2}-\frac{d}{6}) (m_1^2)^{\frac{1}{2}+\frac{d}{6}} \nonumber\\
=\frac{m_1^2}{48 \pi^2}\frac{\Gamma(\frac{\epsilon}{6})}{(1-\frac{\epsilon}{6})}
\frac{M^{\epsilon} (m_1^2)^{-\frac{\epsilon}{6}}}{(4 \pi)^{\frac{\epsilon}{2}}}.
\end{eqnarray}
Expanding the last expression in powers of $ \epsilon$ we get the final result:
\begin{equation}
U_{m_1}=\frac{m_1^2}{48 \pi^2}(\frac{6}{\epsilon}-\gamma-3\ln(4 \pi))
+\frac{m_1^2}{48 \pi^2}(1-\ln(\frac{m_1^2}{M^6})),
\end{equation}
here we have a $\frac{1}{\epsilon}$ divergence corresponding to a logarithmic divergence if
we use a momentum cutoff, $\Lambda$, instead of dimensional regularization.

The fourth term in (\ref{effpot}), that comes from the gauge field, can also be easily integrated with
the previous approach, and we get
\begin{eqnarray}
U_{m_e}=\frac{1}{2}\int\frac{d^3 p}{(2\pi)^3}p^2\sqrt{p^2+m_e^2}
=-\frac{M^\epsilon}{4\sqrt{\pi}}\int_0^\infty d\alpha \alpha^{-3/2}\int\frac{d^d p}{(2\pi)^d}
p^2\exp^{-\alpha(p^2+m_e^2)} \nonumber\\
=-\frac{M^\epsilon}{4\sqrt{\pi}}\frac{1}{(2\pi)^d}\frac{2 \pi^{d/2}}{\Gamma(d/2)}
\frac{\Gamma(1+\frac{d}{2})}{2}(m_e^2)^{\frac{3}{2}+\frac{d}{2}}\Gamma(-\frac{3}{2}-\frac{d}{2}).~~~
\end{eqnarray}
For $d=3-\epsilon$, expanding in powers of $\epsilon$ we get:
\begin{equation}
U_{m_e}=\frac{m_e^6}{128 \pi^2}(\frac{2}{\epsilon}-\gamma-\ln(4 \pi))
+\frac{m_e^6}{128 \pi^2}(\frac{11}{6}-\ln(\frac{m_e^2}{M^2})).~~
\end{equation}

For the last, mixed term in (\ref{effpot}),we proceed as follows:
we rescale with the scalar mass $m_2$ the integration
variables $\omega\rightarrow x=\omega/m_2$ and $p\rightarrow y=p/m_2^{1/3}$ then this term depends on the ratio $\frac{m_e^2}{(m_2^2)^{1/3}}$
\begin{eqnarray}
\frac{1}{2}Tr\ln\left(1+\frac{m_e^2 m_2^2 p^4}{(\omega^2+p^6)(\omega^2+p^6+m_2^2)}\right) \nonumber\\
=2\frac{m_2^2}{(2 \pi)^3} \int_0^\infty \int_0^\infty dx dy
y^2 \ln\left(1+\frac{m_e^2}{(m_2^2)^{1/3}}\frac{y^4}{(x^2+y^6)(x^2+y^6+1)}\right).
\label{mix1}
\end{eqnarray}
If define now the dimensionless parameter $\lambda=\frac{m_e^2}{(m_2^2)^{1/3}}$ and the function
\begin{equation}
F(\lambda)=\int_0^\infty \int_0^\infty dx dy
y^2 \ln\left(1+ \lambda \frac{y^4}{(x^2+y^6)(x^2+y^6+1)}\right),
\end{equation}
then the mixed term (\ref{mix1}) is written as: $2\frac{m_2^2}{(2 \pi)^3}F(\lambda)$.

The integral is finite for every value of the
parameter $\lambda$, for example $F(1)=0.4838$ for a used typical set of the couplings.
The parameter $\lambda$ is in general a function of the scalar field $\phi$.
For the scalar potential (\ref{ss1}) the parameter $\lambda$ is independent of $\phi$ and
equal to $\lambda=(7!)^{1/3} \frac{e^2}{g^{1/3}}$.

Substituting the results in (\ref{effpot}) the dimensionally regularized one-loop effective potential $U_{\rm eff}$
in the modified minimal subtraction ($\overline{MS}$) scheme is written as \cite{peskin}:
\begin{eqnarray}
U_{\rm eff}=U(\phi)+\frac{m_1^2(\phi)}{48 \pi^2}(1-\ln(\frac{m_1^2(\phi)}{M^6}))
+\frac{m_2^2(\phi)}{48 \pi^2}(1-\ln(\frac{m_2^2(\phi)}{M^6})) \nonumber\\
-\frac{4 m_F^2(\phi)}{48 \pi^2}(1-\ln(\frac{m_F^2(\phi)}{M^6}))+
\frac{m_e^6(\phi)}{128 \pi^2}(\frac{11}{6}-\ln(\frac{m_e^2(\phi)}{M^2}))
+2\frac{m_2^2(\phi)}{(2 \pi)^3}F(\lambda).
\label{effpot2}
\end{eqnarray}

For the scalar potential term (\ref{ss1}) that we consider, we have:
\begin{eqnarray}
U_{\rm eff}=\frac{g}{8!}\phi^8 +\frac{g}{6!}\frac{\phi^6}{48 \pi^2}(1-\ln(\frac{g \phi^6}{6! M^6}))
+\frac{g}{7!}\frac{\phi^6}{48 \pi^2}(1-\ln(\frac{g \phi^6)}{7! M^6})) \nonumber\\
-\frac{h^2 \phi^4}{12 \pi^2}(1-\ln(\frac{h^2 \phi^4}{M^6}))+
\frac{e^6 \phi^6}{128 \pi^2}(\frac{11}{6}-\ln(\frac{e^2 \phi^2}{M^2}))
+\frac{g}{7!}\frac{\phi^6}{4 \pi^3}F(\lambda).
\label{effpot3}
\end{eqnarray}
Some words are necessary in order to understand why we choose the scalar potential (\ref{ss1}) to be of the order $\phi^8$.
We observe that the contribution from the gauge field is proportional to $m_e^6\approx e^6 \phi^6$
and negative for large values of the field $\phi$, turning the potential unstable except if
we have a positive contribution of the same order from the other terms. An inspection of the effective potential
shows that the other terms from the scalar loops have exactly the same behavior. The only possibility
that we have to stabilize the potential for large values of $\phi$ is to choose the bare scalar potential
to behave as $\phi^n$ with $n\geq8$ for large values of $\phi$.\\

\section{Results}

We discuss first the behavior of the $U_{\rm eff}$ versus $\phi$ for zero Yukawa coupling, $h=0$. The one loop effective potential is zero and has a local minimum $\phi=0$.
The contribution of the scalar terms and the gauge term is positive for small values of $\phi$ and negative for large values of $\phi$ as we can see from Eq.~(\ref{effpot3}) for the effective potential. The contribution of the mixed term, however, always comes with a positive sign. The final conclusion on the presence of symmetry breaking at one-loop depends on the dimensionless ratio of the scalar and gauge couplings, because if the gauge term turns negative before the scalar terms then it is possible for a second minimum to appear for $\phi\neq 0$.

We plot the effective potential,
for $h=0$, in Fig.~1: the results are shown in terms of the overall dimension one, ultraviolet scale, $M$. The potential is in units of $M^6$ and $\phi$ has mass dimension zero. The coupling constants are $g=\tilde{g} M^6$, $e=\tilde{e} M$ and $h=\tilde{h} M^3$. With fixed $\tilde{g}=0.1$ and $\tilde{h}=0$ the three curves shown correspond to $\tilde{e}=0.68, 0.69$ and $0.70$, from top to bottom. We see the phenomenon of symmetry breaking induced by one-loop effects as the gauge coupling $\tilde{e}$ is increasing beyond a critical value. The critical value for the gauge coupling $\tilde{e}$ depends on the value of the scalar coupling, $\tilde{g}$.

\vspace{0cm}

\begin{figure}[p]
\begin{center}

\includegraphics[scale=1]{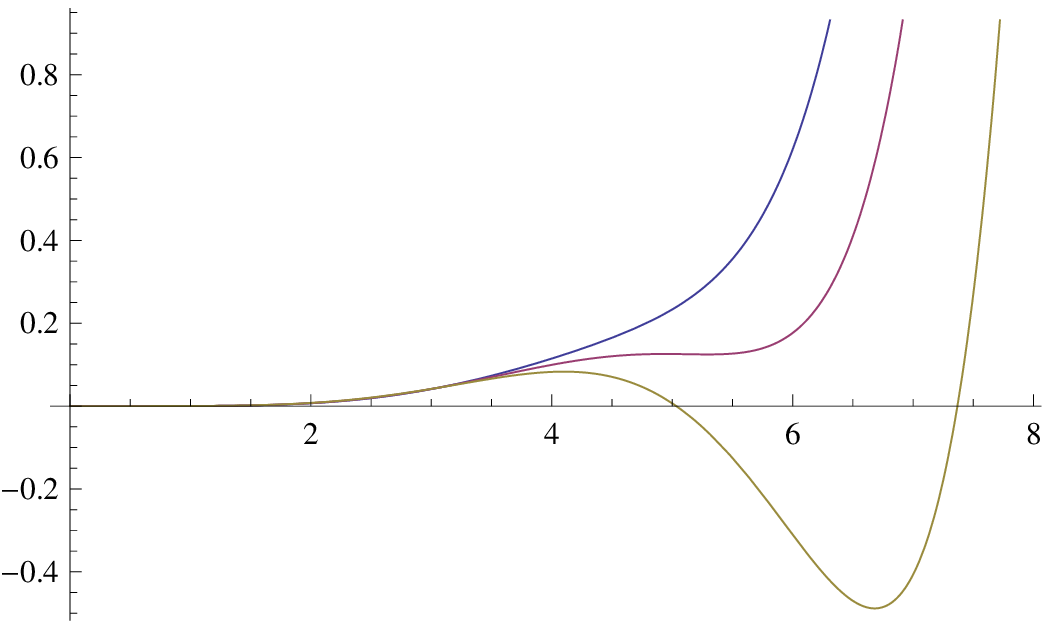}
\caption{ The one-loop effective potential for $h=0$ as a function of $\phi$
 for the scalar-gauge theory.
The potential is plotted in units of $M^6$, $\phi$ is dimensionless.
With fixed $\tilde{g}=0.1$, the three curves shown correspond, from top to bottom, to $\tilde{e}=0.68, 0.69, 0.70$.}

\end{center}
\end{figure}

When we include in the plot the fermion contribution the picture changes, because now the fermion determinant
gives a negative contribution for small values of $\phi$ and turns positive for bigger $\phi$.
We plot the full expression for the effective potential (\ref{effpot3}) with $\tilde{h}\neq 0$ in Fig.~2 and Fig.~3: the results are shown again in terms of the overall ultraviolet scale, $M$. We work with fixed $\tilde{g}=0.1$ and fixed $\tilde{h}=0.2$.

The three curves shown in Fig.~2 correspond to $\tilde{e}=0.15, 0.71, 0.73$, from bottom to top and the figure is drawn for small values of $\phi$. We see that symmetry breaking is induced by one-loop fermion effects for every value of the gauge coupling, $e$. As the gauge coupling increases a second minimum appears for a bigger value of $\phi$ when the gauge coupling is above a critical value. We see this behavior in Fig.~3 for the same values of the gauge couplings, as before, $\tilde{e}=0.15,0.71$ and $0.73$ from top to bottom.
For $\tilde{e}=0.73$ the effective potential has two minima, one for $\phi\approx 1.5$ and the other for $\phi\approx 8.5$. The relevant depth of the two minima is coupling dependent. For $\tilde{e}=0.71$ and $0.15$  the effective potential has only one minimum for $\phi\approx 1.6$
and $\phi\approx 2.2$ correspondingly.\\

\vspace{0cm}

\begin{figure}[p]
\begin{center}

\includegraphics[scale=1]{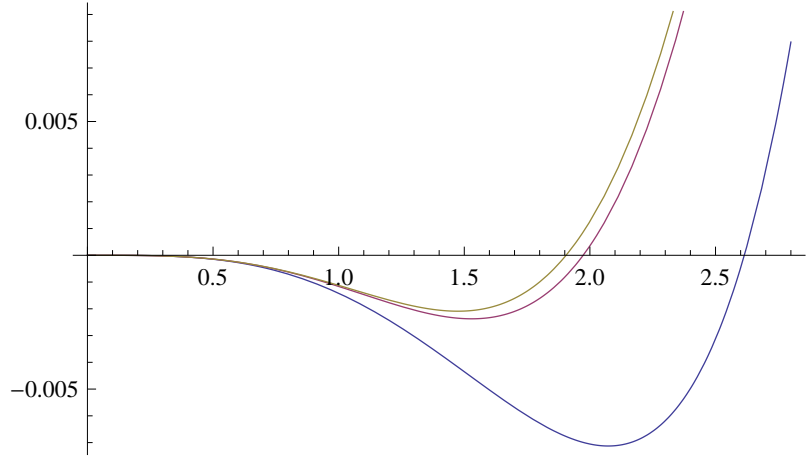}
\caption{ The full one-loop effective potential as a function of $\phi$
 for the scalar-fermion-gauge theory.
The potential is plotted in units of $M^6$ and $\phi$ is dimensionless.
With fixed $\tilde{g}=0.1$ and $\tilde{h}=0.2$, the three curves shown correspond, from bottom to top, to $\tilde{e}=0.15, 0.71, 0.73$.}

\end{center}
\end{figure}

\vspace{0cm}

\begin{figure}[p]
\begin{center}

\includegraphics[scale=1]{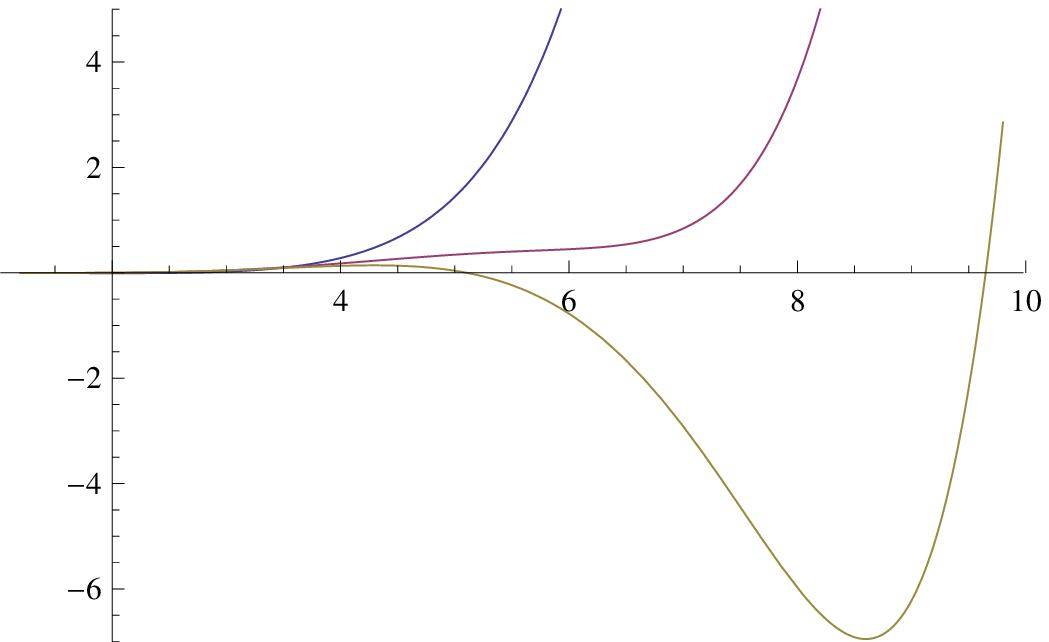}
\caption{ The full one-loop effective potential as a function of $\phi$,
with fixed $\tilde{g}=0.1$ and $\tilde{h}=0.2$, in the region of the second local minimum for $\tilde{e}=0.73$.
The potential is plotted in units of $M^6$ and $\phi$ is dimensionless.
The three curves shown correspond, from top to bottom, to $\tilde{e}=0.15, 0.71$ and $0.73$.
The plot starts from the value $\phi=1.2$.}

\end{center}
\end{figure}

Finally we comment briefly on a Yukawa Lifshitz-type model with a real scalar field \cite{yukawa}.
In \cite{yukawa} a dynamical mass for the fermion appears using the non-perturbative Schwinger-Dyson approach.
The generated fermion mass is non zero when the Yukawa coupling exceeds a critical value.
Using the effective potential techniques as above we can isolate only the fermion contribution.
For the scalar field it is enough to use a mass $\phi^2$ term.
We find then that a symmetry breaking minimum for the potential appears,
for a non zero value of $\phi$, that gives mass to the fermion, for every value of the Yukawa coupling.\\

\textbf{Acknowledgements} We would like to thank N. Vlachos and N. Trakas for useful discussions and D.
Metaxas for reading the manuscript.

\end{document}